\documentclass[12pt]{article}
\usepackage[latin1]{inputenc}
\usepackage[T1]{fontenc}
\usepackage[english]{babel}
\usepackage{wrapfig}
\usepackage{multirow}
\usepackage{graphicx}
\usepackage{amsmath}
\usepackage{here}
\usepackage{amssymb}
\usepackage{mathrsfs}
\usepackage{geometry}
\usepackage[nottoc]{tocbibind}
\usepackage[final]{pdfpages}
\usepackage{hyperref}
\normalsize
\begin{document}
\title{Some Electronic Properties of Metals through q-Deformed Algebras}
\date{}
\maketitle

\begin{center}
Damien Tristant$^{a,b}$, Francisco A. Brito$^{b}$\\
$^{a}$Universit\'e Paul Sabatier de Toulouse, 31062 Toulouse, France\\
$^{b}$Departamento de F\'isica, Universidade Federal de Campina Grande, Caixa Postal 10071, 58109-970 Campina Grande, Para\'iba, Brazil
\end{center}


We study the thermodynamics of metals by applying q-deformed algebras. We shall mainly focus our attention on q-deformed Sommerfeld parameter as a function of q-deformed electronic specific heat. The results revealed that q-deformation acts as a factor of disorder or impurity, modifying the characteristics of a crystalline structure and thereby controlling the number of electrons per unit volume.

\section{Introduction}

All current proposals for quantum groups suggest the idea of classical deformation of an object, which can be, for example, an algebraic group or a Lie group, knowing that the deformed objects lose their group properties. The concept of quantum groups was motivated by problems from a large number of physical situations, ranging from the study of the fractional quantum Hall effect, black holes and high-Tc superconductors, to the non-commutative geometry and quantum theory of super-algebra. This understanding led to ideas that motivated the theory.

In a study conducted in 1904, Frank Hilton Jackson introduces an element called the q-deformed algebra \cite{article_17}. He used a q-analog theorem, which is a generalization involving a new parameter, denoted $q$, having the property to return the original theorem  when selected its limiting case $q\to 1$ \cite{article_18}. 

The motivation of our study lies in the fact that a full understanding of the physical origin of the q-deformation of classical physics is still lacking. It is not clear that there is a standard answer to the q-deformation mechanics inspired by the study of quantum groups. But appeared recently great interest investigating the q-deformed thermodynamic systems at classical level. Deformed theory manages the statistical behavior of complex systems whose underlying dynamics is calibrated on an area of multi-fractal phase governed by the long-range interaction and the effects of long-term memory \cite{article_19}.

A mechanism capable of generating a deformed version of the classical statistical mechanics is to replace the Boltzmann-Gibbs distribution by its deformed version. In this sense, it is postulated a form of entropy that involves a deformed theory of generalized thermodynamics. Thus, some generalizations of statistical mechanics have been proposed \cite{article_20,article_20_2,article_20_3}. In this context, it was shown that a natural realization of q-deformed thermodynamics bosons and fermions can be built on the {\it q-calculus} formalism.

In the recent past, some developing in q-deformed quantum group were studied, for example, analysis properties of a solid throught q-deformed algebra \cite{article_6}.

In fact, a solid consists of a large number of atoms linked by cohesive forces of various kinds. Atomic motion in a solid is very slight, causing every atom to move only within a small neighborhood, and vibrate around its equilibrium point. In a crystalline solid, the equilibrium points of atomic vibrations form a regular spatial structure, such as a cubic or hexagonal structure. Interaction between atoms allows the propagation of elastic waves in solid media, which can be both horizontal and longitudinal. Although this phenomenon is predominant at room temperature, but it is very different at low temperature, where the electronic part take a more important role. Previous studies have analyzed the behavior of the q-deformed phonon contribution \cite{article_6}, but no clear analye has been done about electronic contributions.

In this paper, we add the q-deformed electronic contribution {calculated using the method of free electrons}, discuss new properties and obtained parameters.  The electronic contribution allows us to obtain the q-deformed Sommerfeld parameter. Preliminary results on this parameter show that some metals develop the same characteristics as others through q-deformation. For example, copper (Cu), has the same characteristics as silver (Ag) when $q\approx 0.17$. In addition, we find that the q-deformed Sommerfeld parameter $\gamma_{q}$ is linked to the number of electrons per unit volume, which changes when $q$ varies. As we shall see, this indeed highlights the involvement of impurities reducing the magnitude of the q-parameter. 

The paper is organized as follows. In Sec.~\ref{2} we present the q-deformed algebra. In Sec.~\ref{3} we set up our fermionic system by calculating the q-deformed Fermi-Dirac statistics.  In Sec.~\ref{4} we implement q-deformed Sommerfeld parameter as a function of electronic specific heat and finally, in Sec.~\ref5 we make our final comments.

\clearpage

\section{The q-deformed quantum algebra}
\label{2}

It is necessary to first perform a general construction of the algebra of q-deformed operators.

The q-deformed algebra of the quantum oscillator is defined by q-deformed Heisenberg algebra in terms of creation operator $\hat{a}^{\dag}$, annihilation operator $\hat{a}$ and the quantum number $\hat{N}$,  {by \cite{article_1}, \cite{article_40}}

\begin{equation}
\left[\hat{N},\hat{a}^{\dag}\right]=\hat{a}^{\dag},\,\,\,\,\,\,\,\,\,\,  \left[\hat{N},\hat{a}\right]=-\hat{a},
\end{equation}
and

\begin{equation}
\hat{a}^{\dag}\hat{a}=\left[\hat{N}\right],\,\,\,\,\,\,\,\,\,\,   \hat{a}\hat{a}^{\dag}=\left[1-\hat{N}\right],\,\,\,\,\,\,\,\,\,\, {\hat{a}\hat{a}^{\dag} \mp q\hat{a}^{\dag}\hat{a}=q^{-N}}
\label{eq:Nombres}
\end{equation}

The basic q-deformed quantum number $\left[x\right]$ is defined as \cite{article_2}

\begin{equation}
\left[x\right]\equiv\frac{q^{x}-q^{-x}}{q-q^{-1}},
\end{equation}
where $q$ is an arbitrary real number, $0<q<\infty$, but the formulation is symmetric and can be limited to cases $0<q<1$ or $1<q<\infty$, defined by a symmetry $q\rightarrow q^{-1}$ and the observed value of $q$ has to satisfy the non-additivity property {(see Ref.\cite{article_3} for a comprehensive study on this property in many physical issues)}

\begin{equation}
\left[x+y\right]\neq\left[x\right]+\left[y\right].
\end{equation}

At limit $q\rightarrow 1$, the basic q-deformed quantum number $\left[x\right]$ is reduced to the number $x$ and we find the classical physical properties of materials.
The Pauli exclusion principle is also applicable for the q-deformed fermions, the eigenvalues of the number operator $\hat{N}$ can only be taken the values of $n=0$ and $1$.

The q-Fock space spanned by orthornormalized eigenstates $\left|n\right\rangle$ is constructed according to

\begin{equation}
\left|n\right\rangle=\frac{\left(\hat{a}^{\dag}\right)^n}{\sqrt{\left[n\right]!}}\left|0\right\rangle,\,\,\,\,\,\,\,\,\,\,  a\left|0\right\rangle=0.
\end{equation}

The action of $\hat{a}$, $\hat{a}^{\dag}$ and $\hat{N}$ on the states $\left|n\right\rangle$ in the q-Fock space are known to be

\begin{equation}
\hat{a}\left|n\right\rangle=\sqrt{\left[n\right]}\left|n-1\right\rangle,
\end{equation}

\begin{equation}
\hat{a}^{\dag}\left|n\right\rangle=\sqrt{\left[1+n\right]}\left|n+1\right\rangle,
\end{equation}

\begin{equation}
\hat{N}\left|n\right\rangle=n\left|n\right\rangle.
\label{eq:N_n}
\end{equation}

Having laid the foundation for our q-deformed quantum development, we seek to express the q-deformed Fermi-Dirac statistics.

\clearpage

\section{The q-deformed Fermi-Dirac statistics}
\label{3}

To calculate the mean occupation numbers (q-deformed Fermi-Dirac statistics) of each energy level, we choose the Hamiltonian of non-interacting q-deformed fermions \cite{article_4}

\begin{equation}
\hat{H}=\sum_{\theta}\left(\varepsilon_{\theta}-\mu\right)\hat{N}_{\theta},
\label{eq:hamiltonien}
\end{equation}
where $\hat{N}_{\theta}$ and $\varepsilon_{\theta}$, are respectively the number operator and energy associated with the state label $\theta$, and $\mu$ is the chemical potential of the system.

The main value of the q-deformed occupation number $f_{\theta,q}$ is defined by : 

\begin{equation}
\left[f_{\theta,q}\right]=\frac{1}{\Xi}tr(\exp(-\beta\hat{H})\left[N_{\theta}\right]),
\end{equation}
where $\Xi=tr(\exp(-\beta\hat{H}))$ is the partition function, where $\beta=1/(k_{B}T)$, where $k_{B}$ is the Boltzmann constant. So we find

\begin{equation}
\left[f_{\theta,q}\right]=\frac{1}{tr(\exp(-\beta\hat{H}))}tr(\exp(-\beta\hat{H})a_{\theta}^{\dag}a_{\theta}).
\end{equation}

Thanks to the cyclic properties of the trace \cite{article_5}, and using the above equations (\ref{eq:Nombres}),(\ref{eq:N_n}) and (\ref{eq:hamiltonien}), we can get

\begin{equation}
\frac{\left[f_{\theta,q}\right]}{\left[1-f_{\theta,q}\right]}=\exp\left(-\beta\left(\varepsilon_{\theta}-\mu\right)\right).
\end{equation}

Using the definition of q-deformed number $\left[x\right]$,

\begin{equation}
\frac{\left[f_{\theta,q}\right]}{\left[1-f_{\theta,q}\right]}=\frac{q^{f_{\theta,q}}-q^{-f_{\theta,q}}}{q^{1-f_{\theta,q}}-q^{f_{\theta,q}-1}}=\exp\left(-\beta\left(\varepsilon_{\theta}-\mu\right)\right),
\end{equation}
the final solution is defined by

\begin{equation}
f_{\theta,q}=\frac{1}{2\ln\left(q\right)}\ln\left(\frac{q+\exp\left(\beta\left(\varepsilon_{\theta}-\mu\right)\right)}{q^{-1}+\exp\left(\beta\left(\varepsilon_{\theta}-\mu\right)\right)}\right).
\label{eq:ocupation}
\end{equation}

\begin{figure}[H]
\begin{center}
\includegraphics[width=6.1cm]{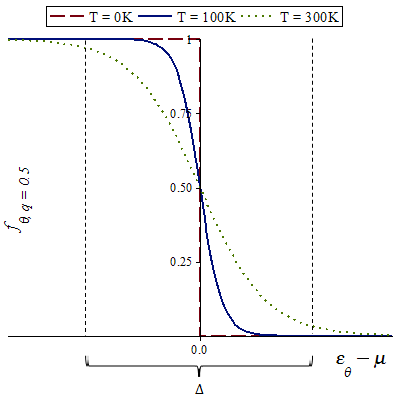}
\,\,\,\,\,\,
\includegraphics[width=7cm]{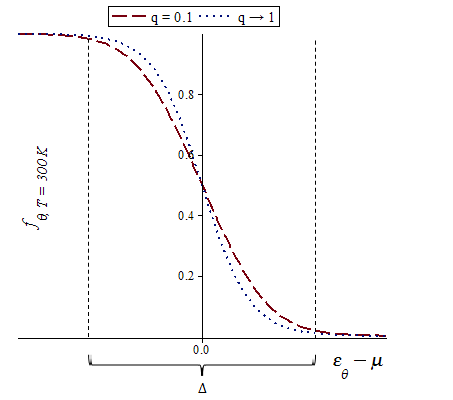}
\end{center}
\caption{We plot the function of q-deformed occupation number (\ref{eq:ocupation}). ({\it Left panel}) For temperatures ($T = 0K$, $100K$ and $300K$) and $q = 0.5$. ({\it Right panel}) For q-parameters ($q = 0.1$ and $1$) and $T=300K$.}
\label{fig:F-D}
\end{figure}

At limit $q\rightarrow 1$, the q-deformed occupation number $\left[f_{\theta,q}\right]$ is reduced to the Fermi-Dirac distribution, $f_{\theta}=1/(\exp\left(\beta\left(\varepsilon_{\theta}-\mu\right)\right)+1)$. We get the same properties as a function of the Fermi-Dirac distribution, i.e., at absolute zero kelvin, the probability is equal one for energies less than the Fermi energy and zero for energies greater than the Fermi energy. Moreover, whatever the value of the temperature, $f_{\theta}=0.5$ when $\varepsilon_{\theta}=\mu$.

On the right Fig.\ref{fig:F-D}, the decrease in the value $q$ reduces the inclination of the slope of the q-deformed Fermi-Dirac function. We can compare this phenomenon in the classical case when the temperature is increased. Moreover, whatever the value of q-parameter, the interval $\Delta$ is proportional to the temperature. On the other hand, from Fig.~\ref{fig:F-D} (right panel) the interval $\Delta$ decreases as $q$ increases and vice-versa. Thus, as a consequence, $T$ decreases as $q$ increases and vice-versa as we shall see later more evidence for this effect.

\clearpage

\section{Implementation of the q-deformation}
\label{4}

We consider an ideal gas of q-fermions confined in a three-dimensional box. The electrons move in a constant effective potential, which results from the interaction of the electron mean made out off all other electrons and ions.

The q-deformed total number of particules and the q-deformed total energy of the system can be, respectively, expressed as

\begin{equation}
N_q\left(T\right)=\int^{\infty}_{0}\mathrm{d}E\,g(E)f_{\theta,q}(E,T),
\end{equation}

\begin{equation}
U_q\left(T\right)=\int^{\infty}_{0}\mathrm{d}E\, g(E)f_{\theta,q}(E,T)E,
\end{equation}
where, in three-dimension, the function of density states $g(E)=C\sqrt{E}$, where $C$ is a constant related to the mass of the electron $m_{e}$.  

These integrals are of the type, $\int^{\infty}_{0}\mathrm{d}E\,h(E)f_{\theta,q}(E,T)$. They can be evaluated by noting that $f_{\theta,q}(E,T)$ is evolving rapidly around $E=\mu$, when $T\ll T_{F}$. Using the integration by parts we find
\begin{equation}
\int^{\infty}_{0}\mathrm{d}E\,h(E)f_{\theta,q}(E,T)=\left[H(E)f_{\theta,q}(E,T)\right]^{\infty}_{0}-\int^{\infty}_{0}\mathrm{d}E\,H(E)\frac{\partial f_{\theta,q}(E,T)}{\partial E},
\label{IPP}
\end{equation}
where
\begin{equation}
H(E)=\int^{E}_{0}h(E) dE.
\label{eq:H(E)=h(E)}
\end{equation}

\begin{figure}[H]
\begin{center}
\includegraphics[width=6.1cm]{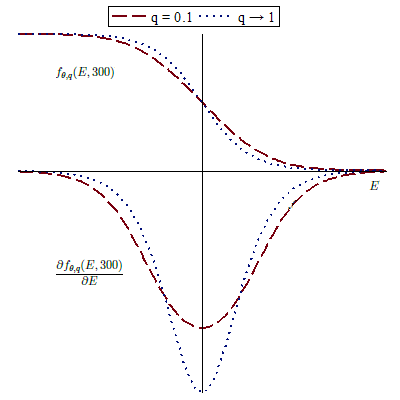}
\end{center}
\caption{We plot the function of q-deformed occupation number (\ref{eq:ocupation}) and its derived function, for q-parameters ($q = 0.1$ and $1$) and $T=300K$, to improve the understanding of the calculations.}
\label{fig:f_df}
\end{figure}

{We note thanks to the Fig.(\ref{fig:f_df}) that $\partial f_{\theta,q}(E,T)/\partial E$ is almost zero everywhere except in the vicinity of $E=\mu$.
The product $H(E)f_{\theta,q}(E,T)$ tends to $0$, at limit $E\rightarrow\infty$. In fact, $f_{\theta,q}(E,T)$ tends to $0$ more rapidly than $H(E)$ tends to infinity, because $H(E)$ is of the form $g(E)$ and $g(E)E$, these functions do not grow exponentially. Moreover, noting that $H(E=0)=0$, and taking into account the fact that $\partial f_{\theta,q}(E,T)/\partial E$ is practically zero for E < 0,  we can write (\ref{IPP}) as}
\begin{equation}
\int^{\infty}_{0}\mathrm{d}E\,h(E)f_{\theta,q}(E,T)=-\int^{\infty}_{0}\mathrm{d}E\,H(E)\frac{\partial f_{\theta,q}(E,T)}{\partial E}.
\label{eq:d(E)}
\end{equation}

If $H(E)$ does not vary too rapidly in the neighborhood of $E=\mu$, it can be developed in Taylor series $H(E)$ and keep the first three terms of the development

\begin{equation}
H(E)=H(\mu)+(E-\mu)\left(\frac{\partial H(E)}{\partial E}\right)_{E=\mu}+\frac{1}{2}\left(E-\mu\right)^{2}\left(\frac{\partial^{2}H(E)}{\partial E^2}\right)_{E=\mu}.
\label{eq:H(E)}
\end{equation}

By substituting the equation (\ref{eq:H(E)}) in the expression (\ref{eq:d(E)}), we get

\begin{multline}
\int^{\infty}_{0}\mathrm{d}E\,h(E)f_{\theta,q}(E,T)=-H\left(\mu\right)\int^{\infty}_{0}\mathrm{d}E\,\frac{\partial f_{\theta,q}(E,T)}{\partial E}\\
-\left(\frac{\partial H(E)}{\partial E}\right)_{E=\mu}\int^{\infty}_{0}\mathrm{d}E\,\left(E-\mu\right)\frac{\partial f_{\theta,q}(E,T)}{\partial E}-\frac{1}{2}\left(\frac{\partial^2 H(E)}{\partial E^2}\right)_{E=\mu}\int^{\infty}_{0}\mathrm{d}E\,\left(E-\mu\right)^{2}\frac{\partial f_{\theta,q}(E,T)}{\partial E}.
\end{multline}

Realizing a change of variables for the function $f_{\theta,q}$, with $x=\beta\left(E-\mu\right)$ and $\mathrm{d}x=\mathrm{d}E\,\beta$. The lower bound of the integrals can be replaced by $-\infty$ at low temperatures, and using the equation (\ref{eq:H(E)=h(E)}) we obtain

\begin{multline}
\int^{\infty}_{0}\mathrm{d}E\,h(E)f_{\theta,q}(E,T)=\mathbf{-1-h(\mu)}\frac{1}{\beta}\int^{\infty}_{-\infty}\mathrm{d}x\,x\frac{\partial f_{x,q}}{\partial x}\\
-\left(\frac{\partial h(E)}{\partial E}\right)_{E=\mu}\frac{1}{2\beta^{2}}\int^{\infty}_{-\infty}\mathrm{d}x\,x^{2}\frac{\partial f_{x,q}}{\partial x},
\label{eq:big}
\end{multline}
where

\begin{equation}
\frac{\partial f_{x,q}}{\partial x}=\frac{1}{2\ln\left(q\right)}\left(\frac{1}{1+q\exp\left({-x}\right)}-\frac{1}{1+q^{-1}\exp\left({-x}\right)}\right).
\label{eq:df}
\end{equation}

Equation (\ref{eq:big}) is composed of three terms. Whatever the value of $q$, in the first term, the integral equals $-1$; in the second term, the integral equals $0$, because the function is odd; only the third term depends on the q-deformation, where the integral is negative.

We put for the continuation of our calculations the integral
\begin{equation}
\int^{\infty}_{-\infty}\mathrm{d}x\,x^{2}\frac{\partial f_{x,q}}{\partial x}=I\left(q\right).
\label{eq:I_q}
\end{equation}

\begin{figure}[H]
\begin{center}
\includegraphics[width=6.5cm]{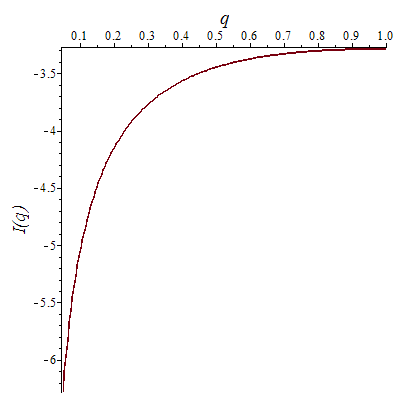}
\end{center}
\caption[Different values of the integral (\ref{eq:I_q}) calculated with some values of $q$.]{The different values of the integral (\ref{eq:I_q}) calculated with some values of $q\in[0,1]$.}
\label{tab:Ik}
\end{figure}

Fig.\ref{tab:Ik} shows some solutions of the integral (\ref{eq:I_q}) for different values of $q$. {In addition, we can see when $q$ approaches $1$, $I(q)$ tends to $-3.2898$}. We use {these informations} later in our calculations.

Now, we can use equation (\ref{eq:big}) to obtain the q-deformed thermodynamic quantities.

\subsection{The q-deformed total number of particles and chemical potential}

The q-deformed number of particules is defined by

\begin{equation}
N_q\left(T\right)=\int^{\infty}_{0}\mathrm{d}E\,g(E)f_{\theta,q}(E,T).
\label{eq:N_q}
\end{equation}

Using the equations (\ref{eq:big}), (\ref{eq:I_q}) and (\ref{eq:N_q}) with $h(E)=C\sqrt{E}$, we get the q-deformed total number of particles $N_q\left(T\right)$, depending of the chemical potential $\mu_q\left(T\right)$

\begin{equation}
N_q\left(T\right)=C\left(\frac{2}{3}\mu_q\left(T\right)^{3/2}-\frac{1}{4\beta^{2}}\mu_q\left(T\right)^{-1/2}I\left(q\right)\right).
\end{equation}

The variation of the chemical potential $\mu_q\left(T\right)$ is obtained by noting that {the q-deformed total number of particles} $N_q\left(T\right)$ is constant when the temperature varies

\begin{equation}
N_q\left(T=0\right)=N_q\left(T\neq0\right).
\end{equation}

We noted previously, when the temperature is zero, the q-deformed Fermi-Dirac statistics $f_{x,q}=1$, so the q-deformed total number of particles $N_q\left(T\right)$ is defined by

\begin{equation}
N_q\left(T=0\right)=\int^{\varepsilon_{F}}_{0}g(E)dE,
\end{equation}
so,

\begin{equation}
\int^{\varepsilon_{F}}_{0}\mathrm{d}E\,\sqrt{E}=\int^{\mu}_{0}\mathrm{d}E\,\sqrt{E}-\frac{1}{2\beta^{2}}{\left(\frac{1}{2\sqrt{\mu}}\right)}I(q),
\end{equation}

\begin{equation}
\int^{\varepsilon_{F}}_{\mu}\mathrm{d}E\,\sqrt{E}=-\frac{1}{2\beta^{2}}{\left(\frac{1}{2\sqrt{\mu}}\right)}I(q).
\end{equation}

The term $\int^{\varepsilon_{F}}_{\mu}\mathrm{d}E\,\sqrt{E}$, takes important values for energies close to the Fermi level $\varepsilon_{F}$. Noting that the Fermi temperature $T_{F}=\varepsilon_{F}/k_{B}$, in the case of free electron gas in three dimensions, the expression of the q-deformed chemical potential is given by

\begin{equation}
\mu_q\left(T\right)=\varepsilon_{F}\left(1+\left(\frac{T}{2T_{F}}\right)^{2}I\left(q\right)\right).
\label{eq:mu}
\end{equation}

Let us note that $\mu_q\left(T\right)$ differs from $\varepsilon_{F}$ by low terms ($T^{2}$). {Then we note that when the temperature tends to zero, the q-deformed chemical potential is equal to the Fermi energy, thus having the same property as the original chemical potential $\mu(T)$ at zero temperature. However, for the temperature not equal to zero, the original chemical potential becomes a q-deformed quantity.
Notice that $\mu$ appears as a limit of integration (expand to first order in $\mu=\varepsilon_{F}$). 
At limit $q\rightarrow 1$, the q-deformed chemical potential $\mu_q\left(T\right)$ is reduced to the classical chemical potential (see, for instance \cite{article_24}, for a recent review)}

\begin{equation}
\mu\left(T\right)=\varepsilon_{F}\left(1-\frac{\pi^2}{12}\left(\frac{T}{T_{F}}\right)^{2}\right).
\end{equation}

Then, one replaces $\mu$ in Eq.(\ref{eq:big}) using Eq.(\ref{eq:mu}), to obtain the expression

\begin{equation}
\int^{\infty}_{0}\mathrm{d}E\,h(E)f_{\theta,q}(E,T)=\int^{\varepsilon_{F}}_{0}\mathrm{d}E\,h(E)-\left(\varepsilon_{F}-\mu\right)h\left(\varepsilon_{F}\right)-\frac{1}{2\beta^{2}}\left(\frac{\partial h(E)}{\partial E}\right)_{E=\varepsilon_{F}}I\left(q\right).
\label{eq:new}
\end{equation}

\subsection {The q-deformed total energy of the system and specific heat of the electron gas}

The q-deformed total energy of the system is defined by

\begin{equation}
U_q\left(T\right)=\int^{\infty}_{0}\mathrm{d}E\,g(E)f_{\theta,q}(E,T)E.
\label{U_q}
\end{equation}

Using the equations (\ref{eq:new}) and (\ref{U_q}) with $h(E)=CE\sqrt{E}$ and the expression of the q-deformed chemical potential (\ref{eq:mu}), we get the q-deformed total energy of the system

\begin{equation}
U_q\left(T\right)=C\varepsilon^{5/2}_{F}\left(\frac{2}{5}-\frac{1}{2}\left(\frac{T}{T_{F}}\right)^{2}I\left(q\right)\right).
\label{U_qfinal}
\end{equation}

We note that the q-deformed total energy of the system depends only on Fermi temperature and q-deformed parameter included in the integral $I\left(q\right)$.

The q-deformed electronic specific heat $C_{Ve_{q}}\left(T\right)$ is defined by

\begin{equation}
C_{Ve_{q}}\left(T\right)=\left(\frac{\partial U_q\left(T\right)}{\partial T}\right)_{V}.
\end{equation}

Differentiating equation (\ref{U_qfinal}), with respect to $T$, we obtain 

\begin{equation}
C_{Ve_{q}}\left(T\right)=-C\varepsilon^{5/2}_{F}\frac{T}{T^{2}_{F}}I\left(q\right).
\label{eq:Cv_q}
\end{equation}
 
But we can simplify this last equation, using the following expressions, $C=(2m_{e}/\hbar^{2})^{3/2}/(2\pi^{2})$, $\varepsilon_{F}=\hbar^{2}k^{2}_{F}/(2m_{e})$ and $T_{F}=\varepsilon_{F}/k_{B}$,
where $k_{F}$ is the Fermi wave vector. Inserting these equations in the expression (\ref{eq:Cv_q}), using $N_{A}$, Avogadro constant and noting $n=N/V=k^{3}_{F}/(3\pi^{2})$ the electronic density, we obtain our final equation of the q-deformed electronic specific heat

\begin{equation}
C_{Ve_{q}}\left(T\right)=-\frac{3}{2}nk_{B}N_{A}\frac{T}{T_{F}}I\left(q\right).
\label{eq:C_qnew}
\end{equation}

At limit $q\rightarrow 1$, the q-deformed electronic specific heat $C_{Ve_{q}}\left(T\right)$ is reduced to the classical electronic specific heat \cite{article_8}

\begin{equation}
C_{Ve}\left(T\right)=\frac{\pi^{2}}{2}nk_{B}N_{A}\frac{T}{T_{F}}.
\end{equation}

We note, that by comparing this result with the specific heat of an ideal gas $3nk_{B}N_{A}/2$, the effect of q-deformed occupation number (whatever the value of $q$) is to reduce the q-deformed electronic specific heat by a factor of $-TI(q)/T_{F}$. At room temperature ($300K$) and whatever the value of $q\in\left[0.1;1\right]$, it is of the order of $\left[10^{-1};10^{-2}\right]$. This explains why we do not detect a significant contribution of the q-deformed electronic specific heat at room temperature. However, it is interesting to study some metals (eg. alloys) at low temperatures, because they provide electrical and thermal properties (eg. superconductivity) different from the room temperature.

From Eq.(\ref{eq:C_qnew}), we can write the q-deformed electronic specific heat as the product of a variable, Sommerfeld parameter $\gamma_{q}$ (depending on the q-deformed parameter) and temperature

\begin{equation}
C_{Ve_{q}}\left(T\right)=\gamma_{q}T,
\label{eq:elec}
\end{equation}
where the Sommerfeld parameter is

\begin{equation}
\gamma_{q}=-\frac{3}{2}\frac{nk_{B}N_{A}}{T_{F}}I\left(q\right).
\label{eq:C_classic}
\end{equation}

The Table 2 of the Ref.\cite{article_10} compares the calculated values of $\gamma_{q\rightarrow 1}$ (free electrons) with experimental values. This comparison requires some foresight. The specific heat of a metal contains two major parts. At room temperature, a solid absorbs heat mainly through the vibrations of ions about their equilibrium positions. However, these contributions vanish as $T^{3}$ at low temperatures. There is a linear contribution (if the value of the q-deformed parameter is fixed) of the specific heat from the electrons. Experimental data support this claim. The author of the reference \cite{article_10} have multiplied the Eq.(\ref{eq:C_classic}) (at limit $q\to1$), by the volume and by the number of conduction electrons $Z$. In addition, this Table 2 of the Ref.\cite{article_10} allows us to verify our calculations, $\gamma_{q\rightarrow 1}$, listed in the Table \ref{tab:2} of this paper. Indeed, when the value of the q-parameter tends to $1$, we find the same {classical} theoretical values for many metals. 

To understand the effects of q-deformation on the Sommerfeld parameter, it is necessary and interesting to plot, for some metals, $\gamma_q$ as a function of $q$. Data is plotted to provide a better view of our results. For illustration purposes, we chose aluminum (Al), copper (Cu), iron (Fe), three materials that can be employed in many areas of interest and bismuth (Bi), gold (Au) and silver (Ag). 

\begin{figure}[H]
\begin{center}
\includegraphics[width=7.5cm]{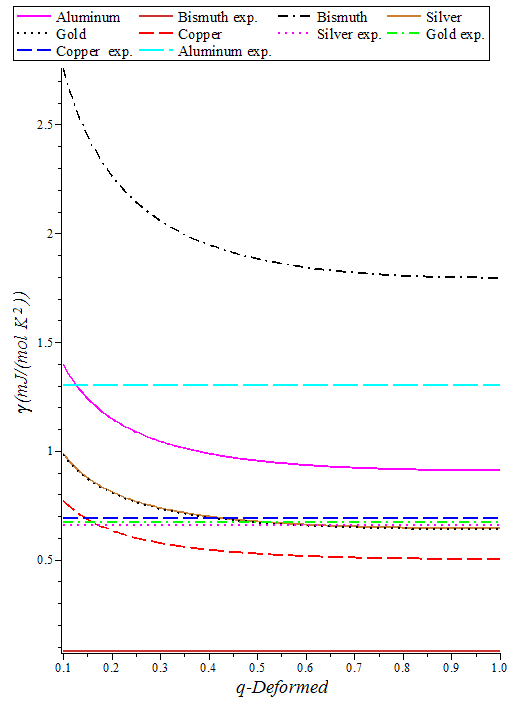}
\end{center}
\caption[The theoretical q-deformed (curves) and experimental (line) Sommerfeld parameter as a function of parameter $0<q<1$ for aluminum, gold, bismuth, silver and copper.]{The theoretical q-deformed (curves) and experimental (line) Sommerfeld parameter as a function of parameter $0<q<1$ for aluminum, gold, bismuth, silver and copper.}
\label{fig:gamma_q1}
\end{figure}

The Fig.\ref{fig:gamma_q1} shows how q-deformation acts on Sommerfeld parameter of these materials. We plot {the curves as} our theoretical data obtained from Eq.(\ref{eq:C_classic}), {whereas the} lines, are experimental data. 

\begin{figure}[H]
\begin{center}
\includegraphics[width=7cm]{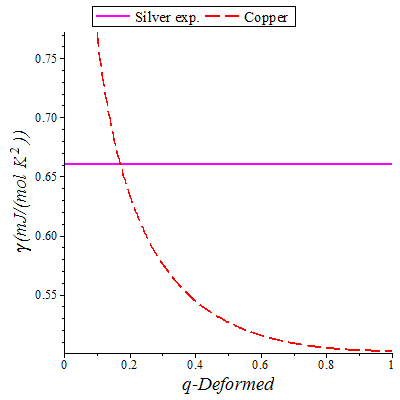}
\end{center}
\caption[The theoretical q-deformed (`dashed' curves) and experimental (line) Sommerfeld parameter as a function of parameter $0<q<1$. On the figure, the experimental value of silver and theoretic values of copper.]{The theoretical q-deformed (`dashed' curves) and experimental (line) Sommerfeld parameter as a function of parameter $0<q<1$. On the figure, the experimental value of silver and theoretic values of copper.}
\label{fig:gamma_q7}
\end{figure}

Let us now define the physical effect caused by the variation of the q-parameter of a material. For this, it is necessary to compare the q-deformed theoretical data of a material with experimental data of another material. In fact, the Fig.\ref{fig:gamma_q7}, shows that Cu reaches Ag experimental value for $q\approx 0.17$. There is a q-parameter value of Cu for which its physical properties are identical to those of Ag ($q\approx 0.17$). Thus, by varying the q-parameter value of a chemical element (eg, $1$ to $0.17$) one modifies the Sommerfeld parameter until achieve the physical properties of another chemical element. This one, amounts to modifying the Fermi energy (depends of the Fermi temperature) and therefore to vary the number of electrons per unit volume $n$. Thus by decreasing the q-parameter, the number of electrons per unit volume move from $n_{Cu}=8.49\cdot10^{22}cm^{-3}$ to $n_{Ag}=5.86\cdot10^{22}cm^{-3}$, given in Ref.\cite{article_8} p.147 (Table 6.1). In addition, we find that the q-deformed Sommerfeld parameter $\gamma_{q}$ is linked to the number of electrons per unit volume, which changes when $q$ varies.

\begin{figure}[H]
\begin{center}
\includegraphics[width=7cm]{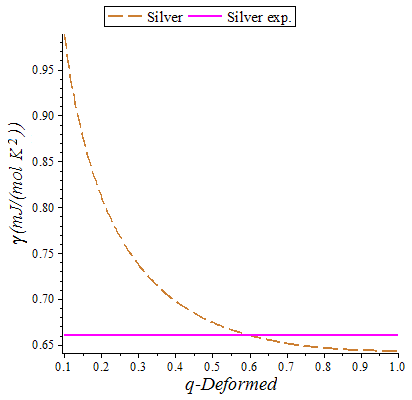}
\,\,\,\,\,\,\,\,\,
\includegraphics[width=7cm]{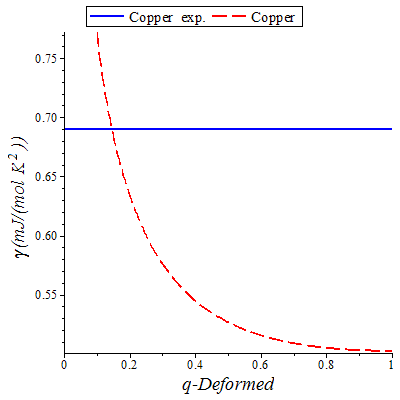}
\end{center}
\caption[The theoretical q-deformed (`dashed' curves) and experimental (line) Sommerfeld parameter as a function of parameter $0<q<1$. On the left figure, the experimental value and theoretic values of silver. On the right figure, the experimental value and the theoretic values of copper]{The theoretical q-deformed (`dashed' curves) and experimental (line) Sommerfeld parameter as a function of parameter $0<q<1$. On the left figure, the experimental and theoretic values of silver. On the right figure, the experimental and theoretic values of copper.}
\label{fig:gamma_q2}
\end{figure}

This is well known, silver is more conductive than copper. Indeed, if we compare the differences between the experimental value $\gamma_{exp.}$ and the theoretical value $\gamma_{q}$ when q tends to 1, Fig.\ref{fig:gamma_q2}, we get respectively for silver and copper, $0.02\, mJ/(mol.K^{2})$ and $0.19\, mJ/(mol.K^{2})$. This one shows that silver is one of the best electrical conductors for which the method of free electrons is appropriate. Indeed, where $q\approx 0.15$, the theoretical value of copper is in agreement with the experimental value.
\\   

\begin{figure}[H]
\begin{center}
\includegraphics[width=7cm]{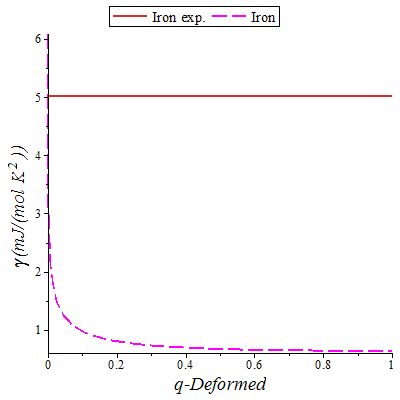}
\end{center}
\caption[The theoretical q-deformed (`dashed' curves) and experimental (line) Sommerfeld parameter as a function of parameter $0<q<1$ for iron]{The theoretical q-deformed (`dashed' curves) and experimental (line) Sommerfeld parameter as a function of parameter $0<q<1$. On the figure, the experimental and theoretic values of iron.}
\label{fig:gamma_q}
\end{figure}

But there are metals in the periodic table to which the estimate of the free electrons of the specific heat is seriously in error, such as iron, Fig.\ref{fig:gamma_q}. According to our results, when $q\approx 2.6\cdot 10^{-4}$, the theoretical value of iron equals the experimental value.

We need to recall the  Kondo's effect \cite{article_23} to understand why there is such a large gap between $\gamma_{q\rightarrow 1}$ and $\gamma_{exp.}$ iron, $4.39\, mJ/(mol.K^{2})$.
Indeed Kondo  assumes the presence of magnetic impurities screened by clouds of electrons. These clouds of electrons have the effect of diffusing the conduction electrons, thereby increasing the resistance.
This behavior turns out to be related to the presence of magnetic impurities in a metal and involves the process where an electron leaves the impurity to be replaced by another electron, which can be of opposite spin (turnaround spin of impurity).
This effect is emphasized through the q-deformation, since the significant intervention of impurities lowers the value of the parameter q-deformed.

\subsection {The q-deformed total specific heat}

As we saw above, the electronic contribution to the q-deformed specific heat is greater than the contribution of phonons at low temperatures. It is well known \cite{article_12}, that the specific heat of a metal $C_{V}$ (electronic and phonons contribution), contains two major parts, low and high temperature relative to the Debye temperature. The same goes for the q-deformed specific heat

\begin{equation}
C_{V_{q}}\left(T\right)=C_{VD_{q}}\left(T\right)+C_{Ve_{q}}\left(T\right).
\label{eq:C}
\end{equation}

The first term in Eq.(\ref{eq:C}) represents the q-deformed Debye specific heat, obtained in Ref.~\cite{article_6}, using the Debye's model. It treats the vibration of the atomic lattice, i.e. phonons in the box

\begin{equation}
C_{VD_{q}}\left(T\right)=3k_{B}N_{A}\left(-\frac{3\left(\frac{\Theta_{D_{q}}}{T}\right)}{\exp\left(\frac{\Theta_{D_{q}}}{T}\right)-1}+\frac{12}{\left(\frac{\Theta_{D_{q}}}{T}\right)^{3}}\int^{\frac{\Theta_{D_{q}}}{T}}_{0}\mathrm{d}\alpha\,\frac{\alpha^{3}}{\exp\left(\alpha\right)-1}\right),
\end{equation}
where $\Theta_{D_{q}}$ is the q-deformed Debye temperature. Table \ref{tab:2} shows some values of $\Theta_{D_{q}}$ for different values of $q$ and different metals \cite{article_6}.
By adding the second term, q-deformed electronic specific heat previously calculated, we obtain the expression of the q-deformed specific heat

\begin{equation}
C_{V_{q}}\left(T\right)=3k_{B}N_{A}\left(-\frac{3\left(\frac{\Theta_{D_{q}}}{T}\right)}{\exp\left(\frac{\Theta_{D_{q}}}{T}\right)-1}+\frac{12}{\left(\frac{\Theta_{D_{q}}}{T}\right)^{3}}\int^{\frac{\Theta_{D_{q}}}{T}}_{0}\mathrm{d}\alpha\,\frac{\alpha^{3}}{\exp\left(\alpha\right)-1}\right)+\gamma_{q}T.
\label{eq:final}
\end{equation}

When the temperature is high in comparison with all the phonon frequencies ($T\gg\Theta_{D_{q}}$), i.e., when all normal mode is in a highly excited state, then the arguments in the exponential are low, the function $C_{VD_{q}}\left(T\right)$ can be expressed in a Taylor series in $\Theta_{D_{q}}/T$

\begin{equation}
C_{VD_{q}}\left(T\right)=3k_{B}N_{A}\left(1-\frac{\left(\frac{\Theta_{D_{q}}}{T}\right)^{2}}{20}\right).
\end{equation}

In addition, we have previously seen, the term containing the Sommerfeld parameter is small at high temperatures. Thus, when the temperature tends to infinity, we obtain the following equation

\begin{equation}
C_{V_{q}}\left(T\right)=3k_{B}N_{A}.
\end{equation}

So we get the Dulong-Petit law, which depends neither on the temperature nor on the q-deformed parameter.
At very low temperatures ($T\ll\Theta_{D_{q}}$), we can write the Eq.(\ref{eq:final}) as

\begin{equation}
C_{V_{q}}\left(T\right)=\frac{12\pi^{4}k_{B}N_{a}}{5}\left(\frac{T}{\Theta_{D_{q}}}\right)^{3}+\gamma_{q}T.
\end{equation}

Thus, as in the usual Debye solid, low temperature q-deformed Debye and the electronic specific heat is proportional respectively to $T^{3}$ and $T$. By studying the q-deformed case at low and high temperatures, we find the same asymptotics of the classical specific heat.

It is useful to have a measure of the temperature at which the specific heat of a metal is no longer dominated by the electronic contribution rather than the contribution of lattice vibrations. Dividing Eq.(\ref{eq:elec}) by the expression at low temperatures of the contribution of phonons, this temperature $T_{0_{q}}$ is obtained

\begin{equation}
T_{0_{q}}=\frac{1}{2\pi^{2}}\sqrt{-\frac{5nI(q)\Theta_{D_{q}}^{3}Z}{2T_{F}}}.
\end{equation}

The Fermi temperature is very large relative to the q-deformed Debye temperature. $T_{0_{q}}$ is typically a few Kelvin. This explains why the linear term in the q-deformed specific heat is observed only at low temperatures. The variation of $T_{0_{q}}$ with the $q$-parameter, e.g. for copper, goes as $T_{0_{q\rightarrow 1}}= 3.22K$, $T_{0_{q=0.5}}= 3.30K$ and $T_{0_{q=0.1}}= 4.00K$. This increasing of the temperature $T_{0_{q}}$ agrees with the observation made previously in Sec.~\ref{3} (the right Fig.\ref{fig:F-D}). Thus, when the q-parameter decreases the temperature $T_{0_{q}}$ increases from a few percent.\\

For the q-deformed case we can observe the changes that occur with Debye temperature, Sommerfeld parameter $\gamma$ and total specific heat. See Table \ref{tb_theta} below for room temperature $T=300K$.

\begin{table}[H]
\centering
\begin{tabular}{|c||c||c|c|c||c|c|c||c|c|c|}
\hline
\multirow{2}*{{\small Element}} & ${\scriptstyle T_{F}^{(a)}}$ & \multicolumn{3}{c||}{${\scriptstyle \Theta_{D_{q}}^{(b)}}$} & \multicolumn{3}{c||}{${\scriptstyle\gamma_{q}^{(c)}}$} & \multicolumn{3}{c|}{${\scriptstyle C_{V_{q}}^{(d)}}$}\\
\cline{3-11}
 & ${\scriptstyle (10^{4})}$ & ${\scriptstyle q=0.1}$ & ${\scriptstyle q=0.5}$ & ${\scriptstyle q\rightarrow 1}$ & ${\scriptstyle q=0.1}$ & ${\scriptstyle q=0.5}$ & ${\scriptstyle q\rightarrow 1}$ & ${\scriptstyle q=0.1}$ & ${\scriptstyle q=0.5}$ & ${\scriptstyle q\rightarrow 1}$ \\
\hline 
 {\small Cs} & {\small 1.76} & {\small 44.5} & {\small 38.6} & {\small 38} & {\small 3.58} & {\small 2.44} & {\small 2.33} & {\small 25.99} & {\small 25.66} & {\small 25.62} \\
\hline
 {\small Rb} & {\small 2.06} & {\small 65.5} & {\small 56.8} & {\small 56} & {\small 3.06} & {\small 2.08} & {\small 1.99} & {\small 25.80} & {\small 25.53} & {\small 25.50} \\
\hline
 {\small K} & {\small 2.37} & {\small 106.5} & {\small 92.4} & {\small 91} & {\small 2.66} & {\small 1.81} & {\small 1.73} & {\small 25.59} & {\small 25.37} & {\small 25.35} \\
\hline
 {\small Pb} & {\small 10.97} & {\small 122.9} & {\small 106.6} & {\small 105} & {\small 2.30} & {\small 1.56} & {\small 1.49} & {\small 25.43} & {\small 25.26} & {\small 25.24} \\
\hline
 {\small Ba} & {\small 4.22} & {\small 128.7} & {\small 111.6} & {\small 110} & {\small 2.98} & {\small 2.03} & {\small 1.94} & {\small 25.61} & {\small 25.38} & {\small 25.36} \\
\hline
 {\small Bi} & {\small 11.43} & {\small 139.3} & {\small 120.8} & {\small 119} & {\small 2.75} & {\small 1.88} & {\small 1.79} & {\small 25.51} & {\small 25.31} & {\small 25.29} \\
\hline
 {\small Na} & {\small 3.66} & {\small 184.9} & {\small 160.4} & {\small 158} & {\small 1.72} & {\small 1.17} & {\small 1.12} & {\small 24.99} & {\small 24.95} & {\small 24.94} \\
\hline
 {\small Au} & {\small 6.42} & {\small 193.1} & {\small 167.4} & {\small 165} & {\small 0.98} & {\small 0.67} & {\small 0.64} & {\small 24.73} & {\small 24.76} & {\small 24.76} \\
\hline
 {\small Sn} & {\small 11.86} & {\small 234.1} & {\small 203} & {\small 200} & {\small 2.12} & {\small 1.45} & {\small 1.38} & {\small 24.84} & {\small 24.82} & {\small 24.81} \\
\hline
 {\small Cd} & {\small 8.66} & {\small 244.6} & {\small 212.1} & {\small 209} & {\small 1.45} & {\small 0.99} & {\small 0.95} & {\small 24.57} & {\small 24.63} & {\small 24.63} \\
\hline
 {\small Ag} & {\small 6.38} & {\small 263.3} & {\small 228.4} & {\small 225} & {\small 0.98} & {\small 0.67} & {\small 0.64} & {\small 24.31} & {\small 24.44} & {\small 24.45} \\
\hline
 {\small Ca} & {\small 5.48} & {\small 269.2} & {\small 233.5} & {\small 230} & {\small 2.30} & {\small 1.57} & {\small 1.49} & {\small 24.66} & {\small 24.68} & {\small 24.68} \\
\hline
 {\small Ga} & {\small 12.11} & {\small 374.5} & {\small 324.8} & {\small 320} & {\small 1.56} & {\small 1.06} & {\small 1.01} & {\small 23.57} & {\small 23.86} & {\small 23.89} \\
\hline
 {\small Zn} & {\small 10.93} & {\small 382.7} & {\small 332} & {\small 327} & {\small 1.15} & {\small 0.78} & {\small 0.75} & {\small 23.37} & {\small 23.72} & {\small 23.75} \\
\hline
 {\small Cu} & {\small 8.17} & {\small 401.4} & {\small 348.2} & {\small 343} & {\small 0.77} & {\small 0.52} & {\small 0.50} & {\small 23.08} & {\small 23.50} & {\small 23.54} \\
\hline
 {\small Li} & {\small 5.43} & {\small 402.6} & {\small 349.2} & {\small 344} & {\small 1.16} & {\small 0.79} & {\small 0.75} & {\small 23.18} & {\small 23.57} & {\small 23.60} \\
\hline
 {\small Al} & {\small 13.53} & {\small 500.1} & {\small 434.4} & {\small 428} & {\small 1.39} & {\small 0.95} & {\small 0.91} & {\small 22.21} & {\small 22.80} & {\small 22.85} \\
\hline
 {\small Fe} & {\small 12.94} & {\small 550} & {\small 477} & {\small 470} & {\small 0.97} & {\small 0.66} & {\small 0.63} & {\small 21.50} & {\small 22.25} & {\small 22.32} \\
\hline
 {\small Be} & {\small 16.67} & {\small 1685.2} & {\small 1461.7} & {\small 1440} & {\small 0.75} & {\small 0.51} & {\small 0.49} & {\small 7.74} & {\small 9.73} & {\small 9.95} \\
\hline 
\end{tabular}
\caption[Chemical elements and their respective Fermi temperature $\,^{a}$ ($K$) \cite{article_8}, Debye temperatures$\,^{b}$ ($K$) \cite{article_6}, Sommerfeld parameter $\gamma_{q}$ $^{c}$ ($\frac{mJ}{mol\cdot K^{2}}$), total specific heat$\,^{d}$ ($\frac{J}{mol\cdot K}$), for $T=300K$ and their deformations for $q=0.1$ and $q=0.5$]{\label{tb_theta}Chemical elements and their respective Fermi temperature $\,^{a}$ ($K$) \cite{article_8}, Debye temperatures$\,^{b}$ ($K$) \cite{article_6}, Sommerfeld parameter $\gamma_{q}$ $^{c}$ ($\frac{mJ}{mol\cdot K^{2}}$), total specific heat$\,^{d}$ ($\frac{J}{mol\cdot K}$), for $T=300K$ and their deformations for $q=0.1$ and $q=0.5$.}
\label{tab:2}
\end{table}

\clearpage

\section{Conclusions}
\label{5}

In this paper, we studied the electronic contribution together with the contribution of phonons \cite{article_6} in the limit of low and high temperatures by applying a q-deformed algebra. The application of this algebra in these well-known problems have resulted in a better understanding of the q-deformation. 

We have obtained the following q-deformed quantities: number of particles, the total energy of the system, the chemical potential and the specific heat. The latter is the more interesting because it allows us to obtain, with simple parameters such as  the Sommerfeld parameter and Debye's temperature, some information about the properties of the q-deformed solid. At limit $q\rightarrow 1$, we find that the results are identical to the classical case referenced in the literature.

In order to extend a previous analysis on thermal and electrical conductivity in q-deformed Debye's solid through the specific heat due to phonons, we also have considered the electronic specific heat. The electronic contribution, allowed us to obtain the q-deformed Sommerfeld parameter. Preliminary results on this parameter show that some metals develop the same characteristics as others through q-deformation. For example, copper (Cu), has the same characteristics as silver (Ag) when $q\approx 0.17$. In addition, we find that this parameter $\gamma_{q}$ is linked to the number of electrons per unit volume, which changes when $q$ varies.

Although some metals do not follow the method of free electrons, for example iron, q-deformation can still give good results. Indeed, this highlights the involvement of impurities reducing the magnitude of the q-parameter. We support this result through the Kondo's theory, which predicts the magnetic impurities intervention in some metals, such as iron.
So, the q-deformation (here formulated with free electrons theory) helps to explain physical phenomenons which are traditionally described by more sophisticated methods such as the Kondo's method.

By adding the electrons contribution to the phonons contribution, we obtain the q-deformed total specific heat. At low temperature, the temperature $T_{0_{q}}$ separating the main contribution of electrons and phonons is gradually shifted to higher temperatures when the q-parameter decreases. For example, for copper (Cu), the temperature increases of $20\%$ if $q=0.1$.

Thus, we see the possibility of applying the q-deformation on different models, with the q-parameter acting as a factor of disorder or impurity. We believe that these defects are mainly due to the factor of doping, i.e., the addition of impurities or defects (donors or acceptors of electrons), which greatly affects the electronic properties of metals. But we still need to analyze other characteristics of metals to establish more precisely the results through the q-deformation.

Our present results where we have mainly addressed the q-deformation on free (conduction) electrons come support and enrich the previous assumptions found in the literature.



\section*{Acknowledgments}

We would like to thank Roland Coratger for stimulating discussions and for supporting this collaboration.  The authors thank University Paul Sabatier of Toulouse (DT) and  CNPq, CAPES, PNPD/PROCAD - CAPES (FAB) for partial financial support.

\end{document}